\documentclass[twocolumn]{aastex63}

\newcommand{{\maxi}}{{\it MAXI}}
\newcommand{{\swift}}{{\it Swift}}
\newcommand{\srcname}{{MAXI J1803$-$298}}

\newcommand{\fscat}{\ensuremath{F_{\rm{scat}}}}

\newcommand{\nh}{\ensuremath{N_\mathrm{H}}}
\newcommand{\ledd}{\ensuremath{L_\mathrm{Edd}}}

\newcommand{\Tin}{T_\mathrm{in}}
\newcommand{\rin}{r_\mathrm{in}}
\newcommand{\Rin}{R_\mathrm{in}}

\newcommand{\msolar}{M_{\odot}}
\newcommand{\mbh}{M_\mathrm{BH}}

\graphicspath{{./}{figures/}}

\shorttitle{Discovery of MAXI J1803$-$298}
\shortauthors{Shidatsu et al.}

\begin{document}

\title{Discovery and Long-term Broadband X-ray monitoring of Galactic Black Hole Candidate MAXI J1803$-$298}

\correspondingauthor{Megumi Shidatsu}
\email{shidatsu.megumi.wr@ehime-u.ac.jp}

\author[0000-0001-8195-6546]{Megumi Shidatsu}
\affil{Department of Physics, Ehime University, 
2-5, Bunkyocho, Matsuyama, Ehime 790-8577, Japan}

\author{Kohei Kobayashi}
\affil{Department of Physics, Nihon University, 1-8-14 Kanda-Surugadai, Chiyoda-ku, Tokyo 101-8308, Japan}
\author{Hitoshi Negoro}
\affil{Department of Physics, Nihon University, 1-8-14 Kanda-Surugadai, Chiyoda-ku, Tokyo 101-8308, Japan}
\author{Wataru Iwakiri}
\affil{Department of Physics, Faculty of Science and Engineering, Chuo University, 1-13-27 Kasuga, Bunkyo-ku, Tokyo 112-8551, Japan}
\author{Satoshi Nakahira}
\affil{Institute of Space and Astronautical Science (ISAS), Japan Aerospace Exploration Agency (JAXA), 3-1-1 Yoshinodai, Chuo, Sagamihara, Kanagawa, 252-5210, Japan}
\author[0000-0001-7821-6715]{Yoshihiro Ueda}
\affil{Department of Astronomy, Kyoto University, Kitashirakawa-Oiwake-cho, Sakyo-ku, Kyoto, Kyoto 606-8502, Japan}
\author{Tatehiro Mihara}
\affil{High Energy Astrophysics Laboratory, RIKEN, 2-1, Hirosawa, Wako, Saitama 351-0198, Japan}
\author{Teruaki Enoto}
\affil{RIKEN Cluster for Pioneering Research, 2-1 Hirosawa, Wako, Saitama 351-0198, Japan}
\author{Keith Gendreau}
\affil{Astrophysics Science Division, NASA Goddard Space Flight Center, Greenbelt, Maryland 20771, USA}
\author{Zaven Arzoumanian}
\affil{Astrophysics Science Division, NASA Goddard Space Flight Center, Greenbelt, Maryland 20771, USA}
\author{John Pope}
\affil{Astrophysics Science Division, NASA Goddard Space Flight Center, Greenbelt, Maryland 20771, USA}
\author{Bruce Trout}
\affil{Microtel, Greenbelt, MD 20770, USA}
\author{Takashi Okajima}
\affil{Astrophysics Science Division, NASA Goddard Space Flight Center, Greenbelt, Maryland 20771, USA}
\author{Yang Soong}
\affil{Astrophysics Science Division, NASA Goddard Space Flight Center, Greenbelt, Maryland 20771, USA}


\begin{abstract}
We report the results from the broad-band X-ray monitoring of the new Galactic black hole candidate \srcname~with the MAXI/GSC and \swift/BAT during its outburst. After the discovery on 2021 May 1, the soft X-ray flux below 10 keV rapidly increased for $\sim 10$ days and then have been gradually decreasing over 5 months. At the brightest phase, the source exhibited the state transition from the low/hard state to the high/soft state via the intermediate state. The broad-band X-ray spectrum during the outburst was well described with a disk blackbody plus its thermal or non-thermal Comptonization. 
Before the transition the source spectrum was described by a thermal Comptonization component with a photon index of $\sim 1.7$ and an electron temperature of $\sim 30$ keV, whereas a strong disk blackbody component was observed after the transition. The spectral properties in these periods are consistent with the low/hard state and the high/soft state, respectively. A sudden flux drop with a few days duration, unassociated with a significant change in the hardness ratio, was found in the intermediate state. 
A possible cause of this variation is that the mass accretion rate rapidly increased 
at the disk transition, which induced a strong Compton-thick outflow and scattered 
out the X-ray flux. Assuming a non-spinning black hole, 
we estimated a black hole mass  of \srcname~as 
$ 5.8 \pm 0.4~(\cos i/\cos 70^\circ)^{-1/2} (D/8~\mathrm{kpc})~M_\sun$ (where $i$ and $D$ are the inclination angle and the distance)  
from the inner disk radius obtained in the high/soft state.

\end{abstract}


\section{Introduction} \label{sec:intro}

Most of the known Galactic black hole X-ray binaries (BHXB) 
exhibit transient behaviors. They are usually dormant in X-rays 
but suddenly enter into an outburst, increasing their X-ray luminosities  
by several orders of magnitude. Because the X-rays are produced 
in the inner parts of the accretion disk, through the release of 
gravitational energy of the accreted gas, the X-ray luminosity 
and energy spectrum depend on the mass accretion rate 
and the structure of the inner disk, and the spacetime in the 
vicinity of the black hole. 
X-ray observations of BHXBs therefore give clues to understanding 
black hole accretion flows and the nature of the black holes themselves. 
In particular, monitoring broadband X-ray spectra during their 
outbursts is helpful to study the evolution of 
the accretion flows over a wide range of mass accretion rates.

\begin{figure*}

\begin{center}
  \includegraphics[width=15cm]{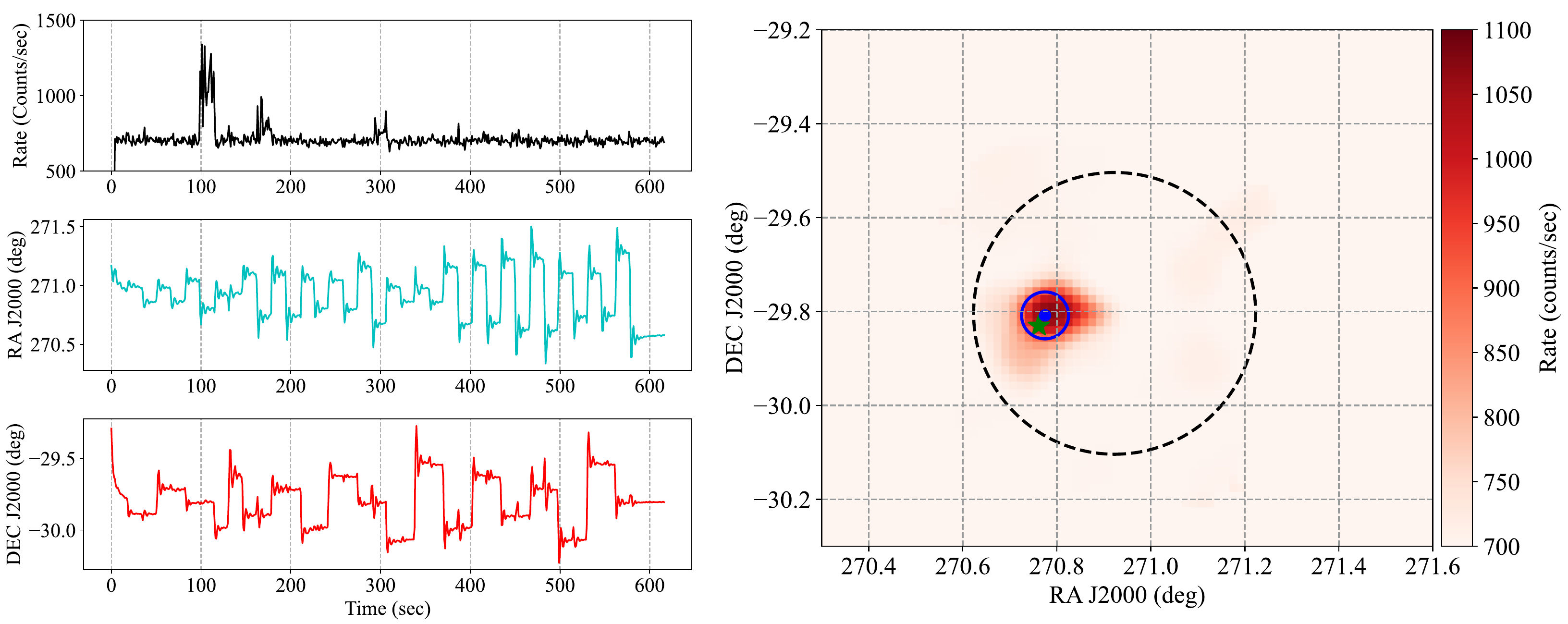}
 \end{center}
\caption{(Left) Time evolution of the NICER raw counts and the pointing direction in R.A. and Dec obtained by the NICER multiple pointings of \srcname. (Right) 2D histogram of the NICER count rate map (see text). The black dotted circle indicates the error circle of the MAXI data reported by \citet{ser21}. The blue filled circle indicates the location of the pixel with the maximum count rate, and the blue solid circle around it represents the FoV of NICER. The green filled star denotes the \srcname~position determined by \swift~ \citep{gro21}.
\label{fig:multipleScan}}
\end{figure*}

The Galactic black hole candidate \srcname~was discovered 
on 2021 May 1 \citep{ser21} with the nova search system 
\citep{neg16} of MAXI \citep{mat09}. Its position 
was then localized by the NICER multiple pointings \citep{gen21} 
and further constrained with \swift~\citep{gro21}. 
Follow-up observations were performed many times in X-rays 
\citep{sgu21, che21, bul21, hom21, xu21, mil21, wan21, 
shi21, jan21, cha21, uba21, ste21, cha21, fen21} 
and other wavelengths \citep{hos21, buk21, esp21, sai21, mat22}.
NuSTAR and NICER found periodic absorption dips, 
and \swift~detected absorption lines likely originating 
in a disk wind \citep{mil21}, both suggestive of a high 
inclination angle above $\sim 70^\circ$. A sign of an outflow 
was also detected in optical spectroscopy \citep{buk21}, where 
p Cygni-like profiles were detected in hydrogen Balmer lines. 

In this article, we report the results from long-term, broadband 
X-ray monitoring of \srcname~over the almost entire outburst, 
using the \maxi/Gas Slit Camera (GSC) and \swift/Burst Alert 
Telescope (BAT) data. 
We used Heasoft version 6.28 and XSPEC version 12.11.1 for data 
reduction and analysis, and adopted the solar abundance table given by 
\citet{wil00}. Throughout the article, errors represent the 90\% 
confidence ranges for one parameter, unless otherwise specified. 

\begin{figure*}[th!]
\epsscale{1.0}
\plotone{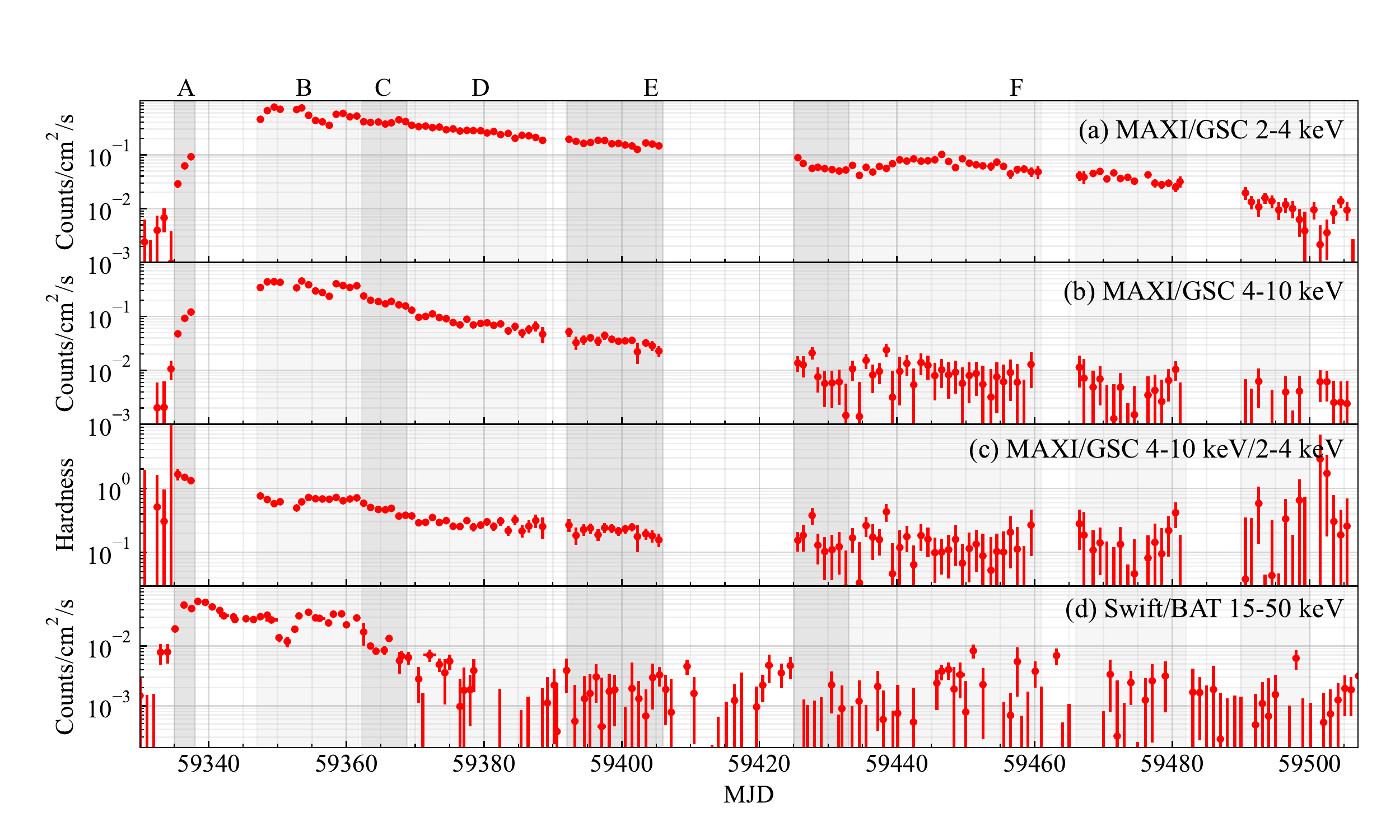}
\caption{2--4 keV and 4--10 keV \maxi/GSC light curves of \srcname, 
their hardness ratio, and the \swift/BAT 15--50 keV light curve,
from top to bottom. All the data points are binned into 1-day intervals.
The error bars indicate 1$\sigma$ statistical errors. 
The phases given in Tab.~\ref{tab:phase} are also indicated. 
\label{fig:LC_longterm}}
\end{figure*}

\begin{figure}[th!]
\plotone{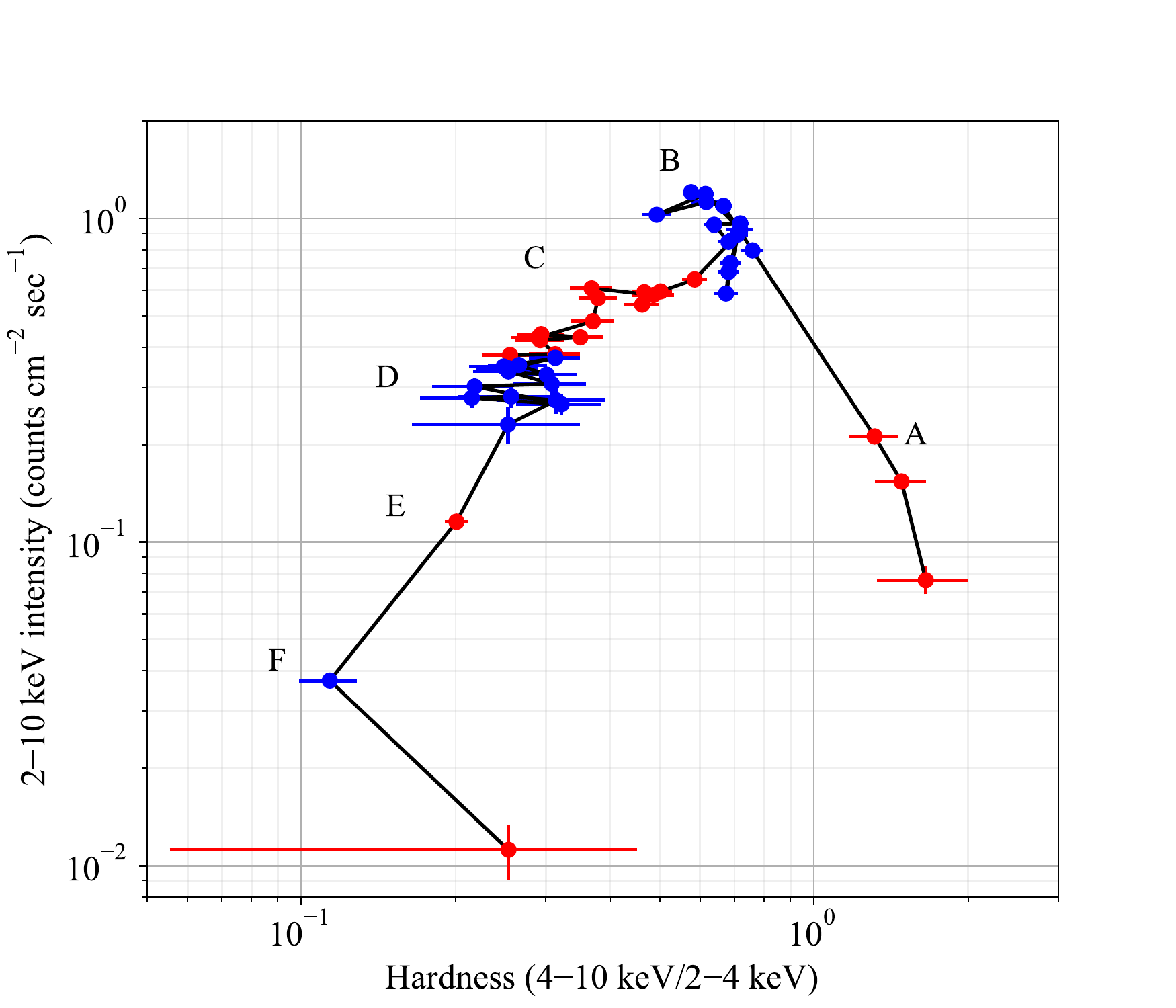}
\caption{Hardness intensity diagram of \srcname, produced 
with the MAXI data in Fig.~\ref{fig:LC_longterm}(a)--(c). 
The phases given in Tab.~\ref{tab:phase} are indicated. 
The data points in Phase E, F, and the later phase 
(MJD 59500--59506) are averaged for each phase, 
to improve statistics. 
\label{fig:HID}}
\end{figure}

\section{Observations and Data Reduction}

\subsection{Localization of the MAXI J1803$-$298 position}
At 19:50 UT on 2021 May 1, the MAXI/GSC nova search system triggered an uncatalogued X-ray transient source. The source position determined by MAXI was $(\alpha^{2000},\delta^{2000}) = (18^h03^m41^s, -29^{\circ} 48^{'}14^{''})$ with a statistical error of about $\pm$0.3 deg and a systematic uncertainty of $\pm$0.1 deg. Following the MAXI alert of a new transient, NICER performed multiple pointings to localize the source position from 03:36 UT on 2021 May 2 using the non-imaging NICER/XTI. Starting from the nominal coordinates reported by MAXI, NICER performed 37 offset pointings with a step interval of 6 arcmin, which is the same as the size of NICER's field of view (FoV). The exposure time for each pointing was about 15 seconds. Figure~\ref{fig:multipleScan} (left) shows the time evolution of the NICER raw counts and the pointing direction coordinates, whereas Figure 1 (right) shows the count rate map created from the raw counts and exposure time map. The vignetting effect in the each pointing was corrected using the vignetting profile of the NICER X-ray concentrator obtained from ray-tracing simulations that considered calibration test results obtained with the 1.5 keV X-rays at the NASA/GSFC 100-m X-ray beamline. The image was smoothed with a Gaussian filter of standard deviation of 1 pixel (1 arcmin). The coordinates of the source estimated from the maximum value of the count rate map is $(\alpha^{2000},\delta^{2000}) = (18^h03^m09^s, -29^{\circ} 48^{'}43^{''})$. After further follow-up observations by \swift~\citep{gro21}, the position of \srcname~was determined to be
$(\alpha^{2000},$ $\delta^{2000}) = 
(18^{\mathrm h}03^{\mathrm m}02^{\mathrm s}.79, 
-29^\circ49'49.''7)$, or ($l$, $b$) = (1$^\circ$.147184, 
-3$^\circ$.727501). This is consistent with the coordinates obtained from the NICER multiple pointing observation within their field of view of 3 arcmin radius.

\subsection{MAXI}
MAXI has been monitoring \srcname~since its discovery 
on 2021 May 1 (MJD 59335). In this work, we used all the  available 
GSC data taken from 2021 April 26 (MJD 59330) to October 19 
(MJD 59506).  
We produced MAXI/GSC long-term light curves through 
a point-spread-function (PSF) fit method \citep{mor10}. 
To extract the GSC spectra, we utilized the MAXI on-demand 
software \citep{nak13} and the Calibration Database (CALDB) 
latest as of 2021 May. The source and background extraction 
regions were defined as a circle with a 1.6$^\circ$ radius and 
an annulus within radii from 1.7$^\circ$ to 3$^\circ$, respectively, 
both centered at the source position. 

Although the source region includes the neighbouring source XTE J1807$-$294 located 
0.96$^\circ$ apart, its contamination is likely to be negligible. To evaluate the 
contamination flux level from XTE J1807$-$294, we  extracted the time-averaged 
spectrum for 1 year before the 
discovery (MJD 58970--59334) from the same source region. However, we did not 
detect source signals significantly, and obtained an upper limit for the 2--10 keV 
flux of 0.8 mCrab ($2.4 \times 10^{-11}$ erg cm$^{-2}$ s$^{-1}$), assuming the Crab 
spectrum ($\Gamma = 2.1$, $\nh = 3 \times 10^{21}$ cm$^{-2}$). This is a negligible 
level in our analysis. To avoid source contamination to the background data, we 
excluded, from the background region, areas within certain radii 
(depending on the source fluxes) around the nearby bright sources: $1.0^\circ$, 
$1.5^\circ$, $1.8^\circ$, $1.4^\circ$, and $2.3^\circ$ 
around SAX J1747.0$-$2853, 1A 1742$-$294, XTE J1751$-$305, 
H 1755$-$338, and GX 5$-$1, respectively, 
all of which are located $3.5^\circ$--5.0$^\circ$ apart from \srcname. 
We determined these exclusion radii based on the 2--10 keV GSC
image in MJD 59335--59350 obtained with the on-demand software, 
so that the PSFs of the sources were sufficiently covered.

\subsection{\swift}

We also used the \swift/BAT survey-mode data for the same period 
as the MAXI/GSC data. The BAT data were downloaded from the HEADAS 
archive\footnote{\url{https://heasarc.gsfc.nasa.gov/FTP/swift/data/obs/}}, 
and were reduced with the \swift/BAT CALDB released on 2017 October 16. 
The data were first processed with the ftool {\tt batsurvey}. The tool 
produces files that list the count rates of the sources in the BAT 
field of view for the individual scans. From these products we 
calculated the 1-day averaged count rates of \srcname~using the 
ftool {\tt ftcalc} and compiled them into a light curve. The 
time-averaged spectrum and the response file for each continuous 
scan were produced with the dedicated script {\tt make\_survey\_pha}
from the {\tt batsurvey} products. 
We created time-averaged spectra for longer time intervals 
by merging the spectra and response files via ftools {\tt mathpha} 
and {\tt addrmf} and used them in the spectral analysis 
(see Sec.~\ref{sec:ana}). In this analysis, we discarded 
the spectral bins when only upper limits on the count rates were 
obtained.

\section{Light Curves and Hardness-intensity Diagram}
\label{sec:lc_hid}

Figure~\ref{fig:LC_longterm} presents the MAXI light curves 
and hardness ratios, and \swift/BAT hard X-ray light curve 
of \srcname. In the initial phase of the outburst, 
the source flux rapidly increased by $\sim$ 2 orders 
of magnitude. The \swift/BAT data suggest that the 
outburst started at least a few days before the 
discovery with the MAXI/GSC on MJD 59335. 
After the source was out of MAXI's field of view from 
MJD 59338 to MJD 59346, it reached the peak with $\sim 0.6$ 
Crab in 2--4 keV and $\sim 0.4$ Crab in 4--10 keV. 
The hardness ratio dropped in this data gap, suggesting 
that the source started the state transition from the 
low/hard state to the high/soft state, as reported by \citet{shi21}. 
The \swift/BAT hard X-ray flux reached its peak around 
MJD 59339, $\sim 10$ days before the flux peak below 10 keV. 
After the flux peaks, the source has gradually dimmed over 
5 months, with a slight rebrightening around MJD 59440. 
Using the NICER data, \citet{ste21} reported 
that the source returned to the low/hard state between 
MJD 59500 and MJD 59506, although this transition is not very clear 
in Fig.~\ref{fig:LC_longterm} due to low statistics.

\begin{figure}[th!]
\plotone{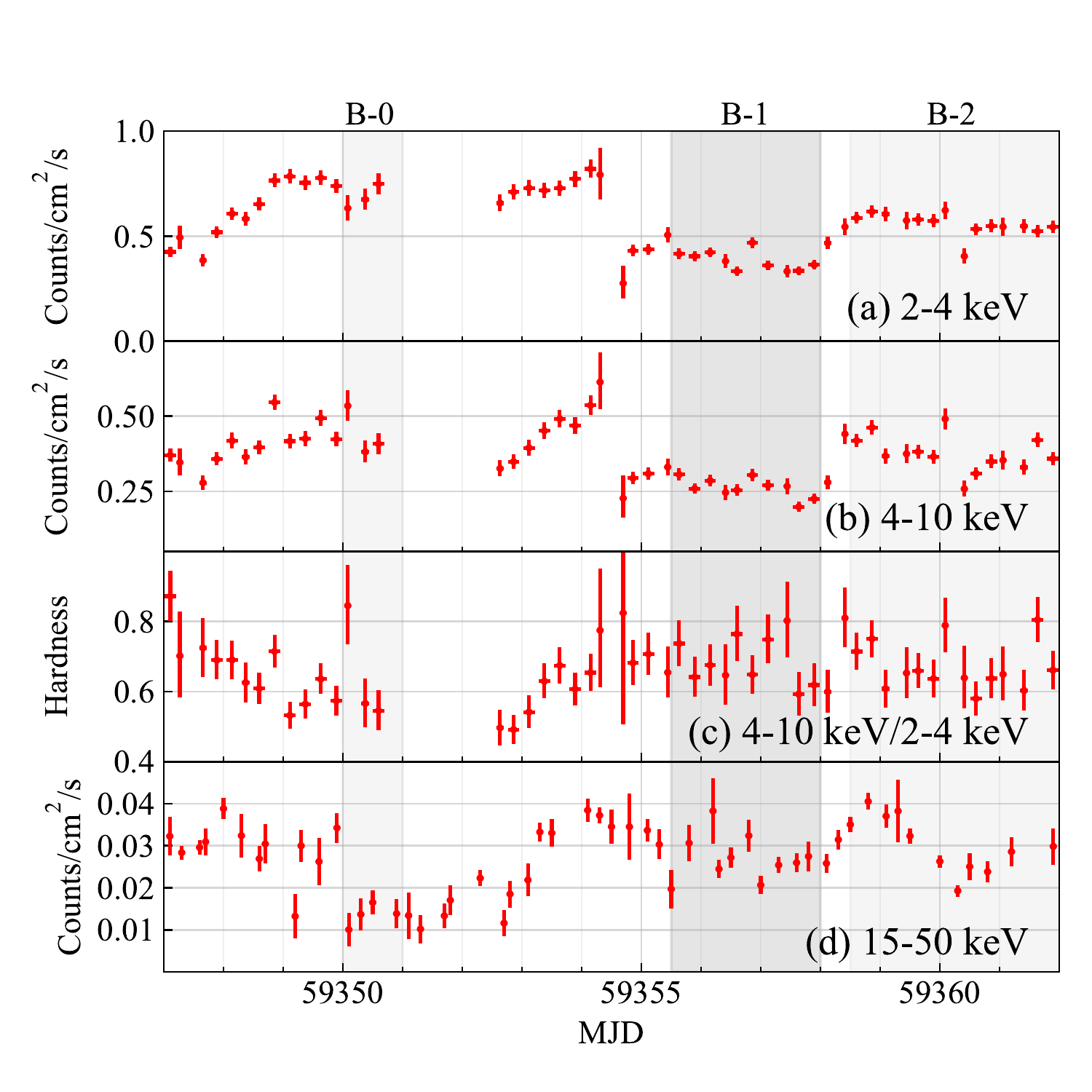}
\caption{Same as Fig.~\ref{fig:LC_longterm}, but 
with 6-hour bins and limited in MJD 59345--59362
(Phase-B). The shadowed regions indicate 
the phases of the highest soft X-ray flux with relatively
weak hard X-rays (Phase B-0), the flux drop in the 
soft X-rays (Phase B-1), and after the drop (Phase B-2),
used in Section~\ref{subsec:spec_var}.
\label{fig:LC_4orb}}
\end{figure}

\begin{deluxetable}{ccccc}[ht!]
\tablecaption{List of phases. \label{tab:phase}}
\tablecolumns{5}
\tablenum{1}
\tablewidth{0pt}
\tablehead{
\colhead{ID} & \colhead{State} & \colhead{Period} & \multicolumn{2}{c}{Exposure} \\
\colhead{} & \colhead{} & \colhead{(MJD)} & \colhead{GSC} (cm$^2$ ks) & \colhead{BAT} (ks)}
\startdata
A & low/hard & 59335--59337 & 7.9 & 22 \\
B & intermediate & 59347--59361 & 31.9 & 175 \\
C & high/soft & 59362--59375  & 20.3 & 109 \\
D & high/soft & 59376--59389 & 16.6 & 98 \\
E & high/soft & 59393--59432 & 36.8 & 400 \\
F & high/soft & 59433--59500 & 89.7 & 657 \\
\enddata
\end{deluxetable}

In Figure~\ref{fig:HID} we plot the hardness intensity diagram,  
where a counter-clockwise path is seen like other transient 
BHXBs. On the basis of the behavior of the source fluxes and 
hardness ratio, we defined six phases listed in Table~\ref{tab:phase} 
and used them in the following spectral analysis. Here, the data gaps 
of MAXI/GSC were omitted. The data before the discovery  
(until MJD 59334) and after MJD 59500 were also ignored, 
because we were unable to obtain 
statistically meaningful spectra due to the too low flux. 
Considering the hardness intensity diagram, 
the source was likely to be in the low/hard state and the high/soft 
state in Phases A and C--F, respectively, in which the source had 
a high and low hardness ratios. In Phase B, the hardness 
ratio was between the values in Phases A and C--F, indicating that 
the source was likely to be in the intermediate state. These are 
confirmed in the following spectral analysis. 

We also created 6-hour bin light curves (Figure~\ref{fig:LC_4orb}) 
in Phase B to study variability during the intermediate state. 
Interestingly, the source exhibited a flux drop by $\sim 40$\%
during MJD 59355--59358 without changing the level of the 
hardness ratio significantly. In addition, the \swift/BAT 
hard X-ray flux was dropped by $60$\% around the soft 
X-ray peak (MJD 59350).

\begin{figure*}[ht!]
\epsscale{1.1}
\plotone{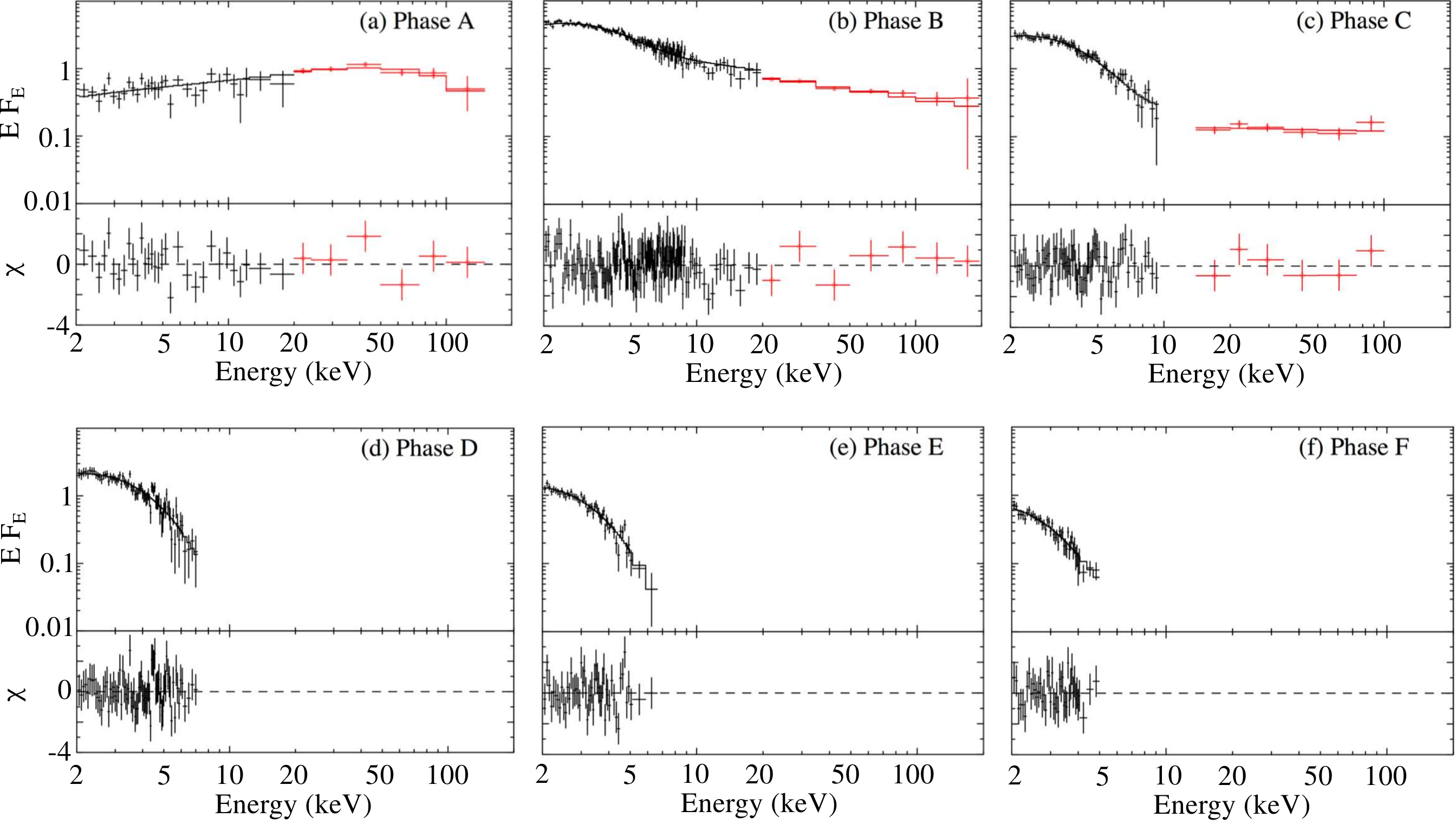}
\caption{(Top) unfolded \maxi/GSC (black) and \swift/BAT (red open square) 
spectra for the individual phases given in Tab.~\ref{tab:phase}, fitted with 
the{\tt TBabs*nthcomp} model (for Phase A) or {\tt TBabs*simpl*diskbb} model 
(Phase B--F). The units of the ordinate axes are keV (Photons cm$^{-2}$ 
s$^{-1}$ keV$^{-1}$).
(Bottom) Residuals for the best-fit models.
\label{fig:MAXIBATspec}}
\end{figure*}

\section{Spectral Analysis and Results}
\label{sec:ana}

\begin{deluxetable*}{cCCCCCCCC}[ht!]
\tablecaption{Best-fit 
spectral parameters\tablenotemark{a} in each phase. \label{tab:fit}}
\tablecolumns{9}
\tablenum{2}
\tablewidth{0pt}
\tablehead{
\colhead{Phase} & \colhead{$\Gamma$} & \colhead{$kT_\mathrm{e}$} & \colhead{$F_\mathrm{scat}$} &
\colhead{$kT_\mathrm{in}$} & \colhead{$r_\mathrm{in}$\tablenotemark{b}} 
& \mathrm{cross~norm.} & \colhead{$\chi^2/\mathrm{dof}$} & \colhead{$F_X$\tablenotemark{c}} \\
\colhead{} &
\colhead{} &
\colhead{keV} &
\colhead{} &
\colhead{keV} &
\colhead{km} &
\colhead{BAT/GSC} &
\colhead{} &
\colhead{$10^{-9}$ erg s$^{-1}$ cm$^{-2}$}
}
\startdata
A & 1.7 \pm 0.1 & 26^{+18}_{-7} & - 
& 0.1~\mathrm{(fixed)} & - & 1.0^{+0.3}_{-0.2} & 38/35 & 5.5^{+0.6}_{-0.5} \\
B & 2.46^{+0.08}_{-0.07} & - & 0.22^{+0.03}_{-0.02} 
& 0.88 \pm 0.03 & 46^{+4}_{-3} & 0.8 \pm 0.1 & 155/154 & 19.8 \pm 0.5 \\
C & 2.1 \pm 0.2 & - & 0.04 \pm 0.02  
& 0.86 \pm 0.04 & 40 \pm 4 & 0.5^{+0.4}_{-0.2} & 65/77 & 11.7^{+0.8}_{-0.7} \\
D & 2.1 \mathrm{(fixed)} & - & <0.03 
& 0.81^{+0.03}_{-0.06} & 38^{+7}_{-3} & -& 77/75 & 7.5^{+0.9}_{-0.4} \\
E & 2.1 \mathrm{(fixed)} & - & <0.02 
& 0.64^{+0.02}_{-0.03} & 50^{+7}_{-3} & - & 52/49 & 4.8^{+0.7}_{-0.2} \\
F & 2.1 \mathrm{(fixed)} & - & <0.04
& 0.56^{+0.04}_{-0.07} & 42^{+17}_{-6} & - & 30/41 & 2.1^{+0.6}_{-0.3} \\ \hline
B-1 & 2.7 \pm 0.2 & - & >0.3
& <0.72 & 93^{+930}_{-34} & 1.1 \pm 0.3 & 46/33 & 17.5^{+7.7}_{-1.9} \\ 
B-2 & 2.5 \pm 0.1 & - & 0.26^{+0.05}_{-0.04} 
& 0.85 \pm 0.06 & 51^{+8}_{-6} & 0.8 \pm 0.1 & 38/35 & 21.7^{+1.0}_{-0.9} \\ \hline
\multicolumn{9}{l}{(Simultaneous fit)} \\
C & 2.1 \pm 0.2\tablenotemark{d} & - & 0.05^{+0.02}_{-0.01} 
& 0.83 \pm 0.02 & 44 \pm 3\tablenotemark{d} & 0.5^{+0.2}_{-0.1} & 234/245\tablenotemark{e} & 11.7^{+0.8}_{-0.7}\\ 
D & \mathrm{(linked)} & - & 0.019^{+0.019}_{-0.018} 
& 0.77 \pm 0.02 & \mathrm{(linked)} & - & ... & 7.5^{+1.0}_{-0.4} \\ 
E & \mathrm{(linked)} & - & <5\times 10^{-3}
& 0.67^{+0.02}_{-0.01}  & \mathrm{(linked)} & - & ... & 5.0^{+0.4}_{-0.3} \\
F & \mathrm{(linked)} & - & <3 \times 10^{-2} 
& 0.56^{+0.01}_{-0.02}  & \mathrm{(linked)} & - & ... & 2.2^{+0.6}_{-0.3} \\ 
\enddata
\tablenotetext{a}{The {\tt TBabs*nthcomp} model was applied for Phase A, while {\tt TBabs*simpl*diskbb} was adopted to the other phases. The $N_\mathrm{H}$ 
value of {\tt TBabs} was fixed at $3.0\times 10^{21}$ cm$^{-2}$.}
\tablenotetext{b}{Inner disk radius estimated from the relation $r_\mathrm{in} = \sqrt{N_\mathrm{dbb}/\cos i} \, (D/10~\mathrm{kpc})$, where $N_\mathrm{dbb}$, $D$, and $i$ are distance and 
inclination angle. respectively. $i=70^\circ$ and $D=8$ kpc were assumed here.
}
\tablenotetext{c}{Unabsorbed 0.01--100 keV flux.}
\tablenotetext{d}{The normalizations of {\tt diskbb} and $\Gamma$ 
of {\tt simpl} were linked among the four phases.}
\tablenotetext{e}{Total $\chi^2$/d.o.f. value of all the four spectra.}
\end{deluxetable*}

\subsection{Time-averaged Spectra in the individual phases}
\label{subsec:spec_phase}
Figure~\ref{fig:MAXIBATspec} shows the time-averaged MAXI/GSC and 
\swift/BAT spectra in the individual phases. For the MAXI/GSC 
data, we discarded the spectral bins at high energies when the 
source count rate is below $\sim 10$ \% of the background level. 
In Appendix~\ref{appsec:MAXIspec_folded}, we show the MAXI/GSC 
response folded spectra and 
their background contribution. The \swift/BAT data for Phase 
D, E, and F, was unavailable. In these phases, all of the 
spectral bins gave only upper limits on the count rates.
In Phase A, the source exhibited a hard, power-law shaped 
spectrum with a cutoff at $\sim 50$ keV. 
Then, in Phase B, a strong thermal component, likely originated in the 
standard disk, emerged in the soft X-ray band. Meanwhile, 
the spectral cutoff disappeared in the hard X-ray band. 
A strong, steep power-law shaped component, with a photon index of $\sim 2.5$, was seen in the Phase B spectrum, while in later phases the hard 
X-ray component became weak and harder. 

We analyzed the broad-band X-ray spectra of \srcname~using a 
standard model for BHXBs: the multi-color disk blackbody 
emission \citep[{{\tt diskbb}};][]{mit84} and its Comptonization.
The {\tt diskbb} model is parametrized by the disk temperature 
and the normalization that is determined by the inner disk 
radius $\rin$, the distance $D$, and the inclination angle $i$.
For the Comptonization component, we adopted {\tt nthcomp} for Phase 
A \citep{zdz96,zyc99} and {\tt simpl} \citep{ste09} for Phase B--F. 
The {\tt nthcomp} model calculates a thermally Comptonized 
spectrum using the photon index $\Gamma$, the electron temperature 
$kT_\mathrm{e}$, and seed photon temperature (the inner disk 
temperature $k\Tin$ when the seed spectrum is set to be 
a disk blackbody).
Because the direct disk component is not seen in the soft X-ray band, 
the {\tt diskbb} model was not used and $k\Tin$ was fixed at 
0.1 keV for the Phase-A spectrum. 
The {\tt simpl} model is a convolution model redistributing a fraction ($F_\mathrm{scat}$) of the input seed photons into a power-law profile
with a photon index $\Gamma$. When using this model, we extended 
the energy range used in the calculation down to 0.01 keV and 
up to 1000 keV so that the spectral fit is not affected by 
uncertainties at the upper and lower boundaries of energy. 

We also used the {\tt TBabs} model to account for 
the interstellar absorption. 
Since the MAXI/GSC has sensitivity only above 2 keV, it is 
difficult to determine the column density. The column density 
was therefore fixed at $\nh = 3 \times 10^{21}$ cm$^{-2}$, 
which was obtained in a NICER observation \citep{hom21}. 
We have confirmed that the uncertainty in $\nh$ gives 
only a small effect to the fit parameters; 
they do not change beyond their 90\% 
error ranges when $\nh = 2 \times 10^{21}$ cm$^{-2}$ and 
$\nh = 4 \times 10^{21}$ cm$^{-2}$ were adopted. 
The cross-normalization factor of the \swift/BAT data 
with respect to the MAXI/GSC data was varied to 
account for the uncertainty in the instrumental 
cross calibration and that caused by flux variation. 

\citet{hom21}, \citet{xu21}, and \citet{jan21} 
reported that the source showed strong absorption dips, 
with a $\sim 7$-hour interval and a $\sim 5000$ s duration, 
in which the X-ray intensity was reduced by $\sim$ 50--100\%. 
However, we found that the effect in our spectral analysis 
was negligible. The conclusion of the analysis did not 
change when the data in the dip phases (assuming the above 
interval and duration, and the center of a dip reported by 
\citealt{hom21}) were excluded. This is likely because 
the strong dips were short-lived compared with the data periods 
that we employed. Actually, previous studies of other 
transient BHXBs found that dips were seen only a 
limited period of outbursts \citep[e.g.,][]{kuu13}.

As shown in Fig.~\ref{fig:MAXIBATspec}, the {\tt TBabs*nthcomp} 
model and {\tt TBabs*simpl*diskbb} 
model were able to reproduce the observed spectra. 
The best-fit parameters in each phase are summarized 
in Table~\ref{tab:fit}. The phase C--F spectra were 
dominated by the {\tt diskbb} component in the soft 
X-ray band below 10 keV.
In Phase D--F, the photon index $\Gamma$ was not constrained at all, 
because the hard tail was not clearly observed due to 
the lack of the hard X-ray data. For these phases, 
we adopted the best-fit value in phase C, $\Gamma =2.1$. 
Many previous works of BHXBs indicated that the inner disk 
radius remains constant when the X-ray spectrum 
is dominated by disk blackbody emission \citep[e.g.,][]{ebi93, 
kub04, ste10, shi11}. 
Considering this, we attempted to fit the Phase C--F 
spectra simultaneously, linking the {\tt diskbb} 
normalizations of all the phases. In this fit, $\Gamma$ was 
also linked to one another. As shown in Tab.~\ref{tab:fit}, 
we obtained an acceptable fit, which gave $\rin = 44 \pm 3$ 
($\cos i/\cos 70^\circ$)$^{-1/2}$ ($D$/8~kpc) km.

\subsection{Short-term Variation in Phase B}
\label{subsec:spec_var}
\begin{figure}[ht!]
\plotone{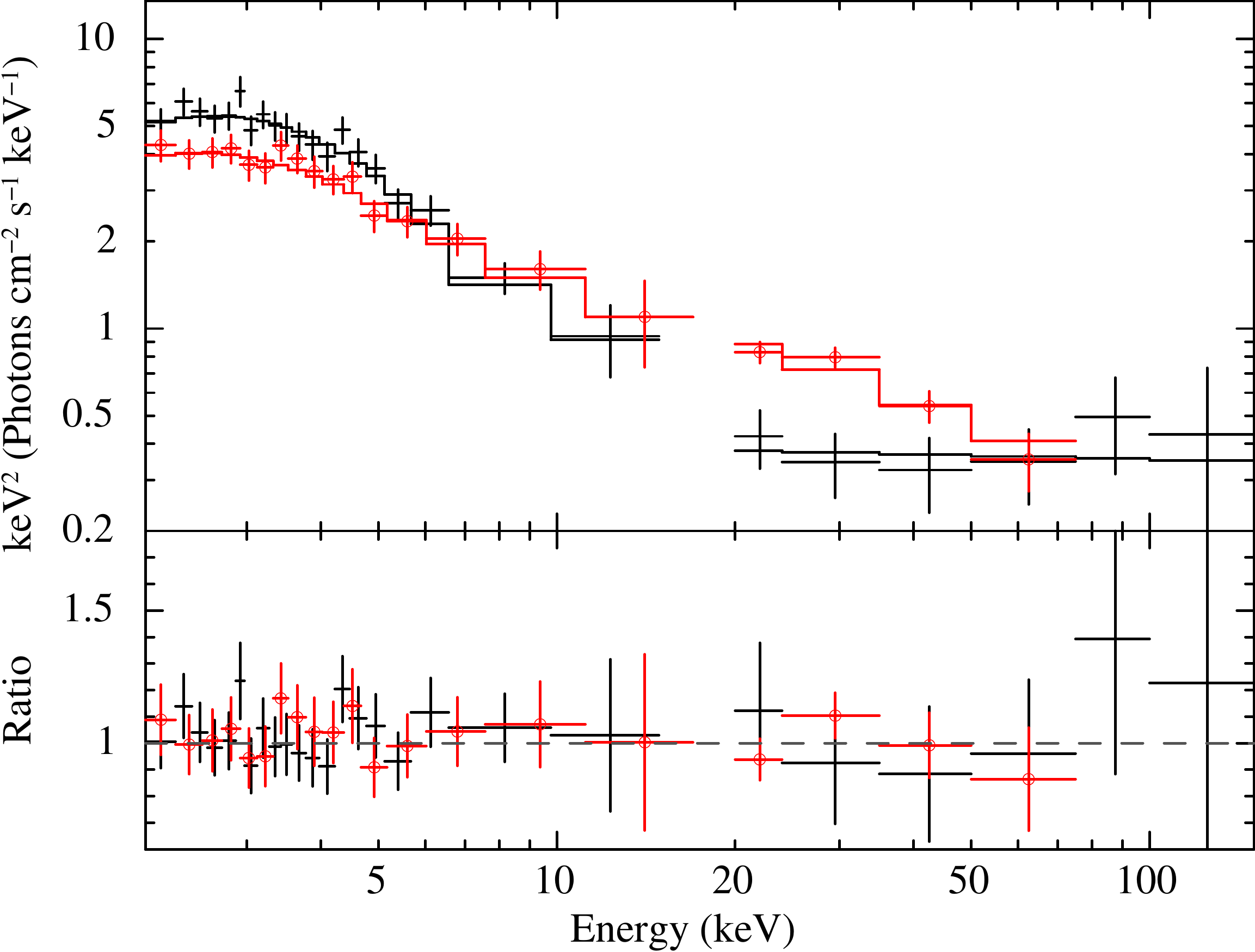}
\caption{One-day-averaged unfolded spectra taken on 
MJD 59347 (red open circles) and MJD 59350 (Phase B-0; black crosses) 
and their best-fit {\tt TBabs*simpl*diskbb} models (top)
and the data versus model ratio (bottom). 
\label{fig:spec_1d}}
\end{figure}

\begin{figure}[ht!]
\plotone{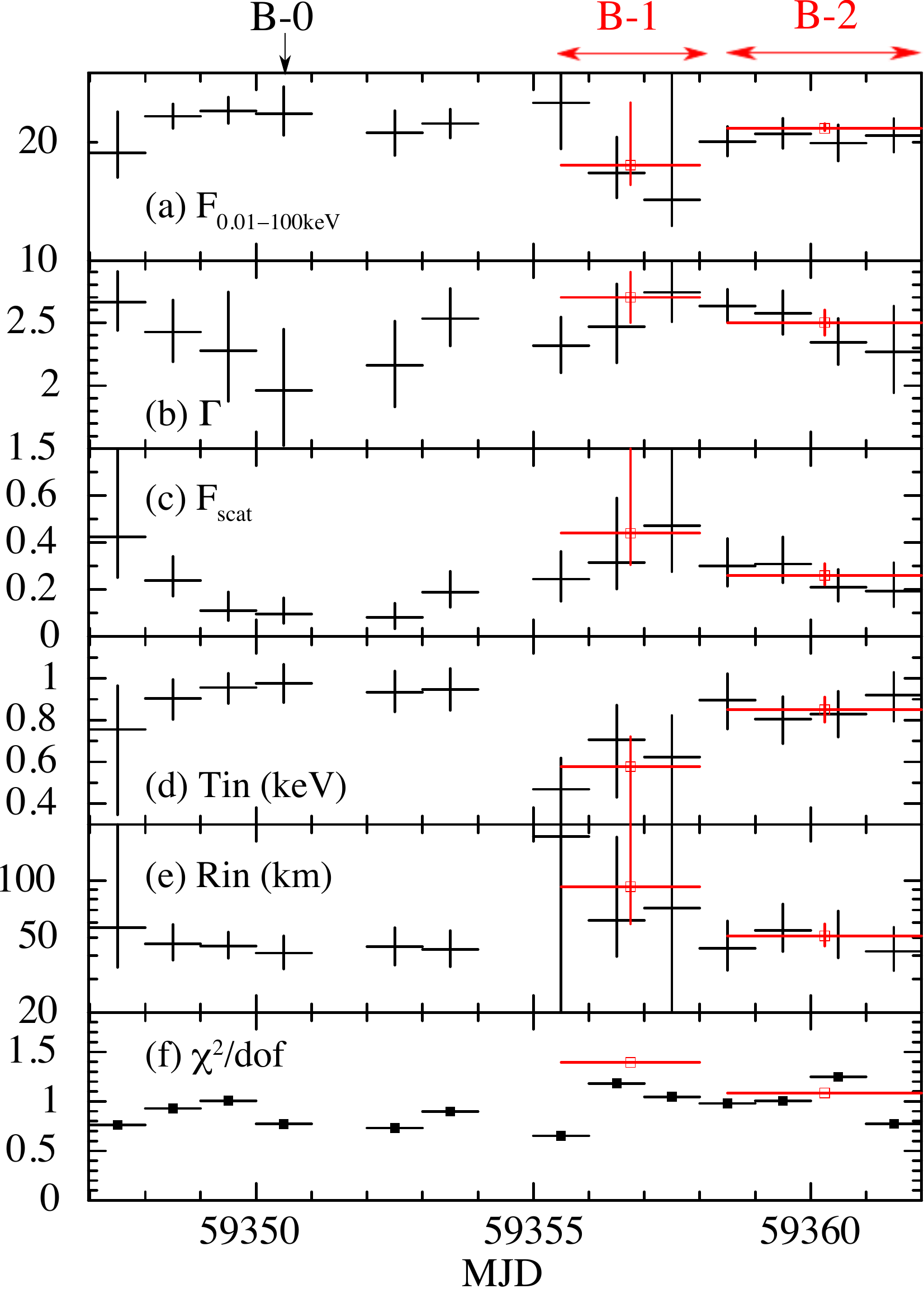}
\caption{Time variations of the best-fit spectral parameters in the intermediate state (Phase B). The top panel plots the unabsorbed 0.01--100 keV flux in units of $10^{-9}$ erg cm$^{-2}$ s$^{-1}$. Results from the 1-day averaged spectra are shown in black and 
those from the Phase B-1 and B-2 spectra in red (with open squares). 
\label{fig:trend_fitpars}}
\end{figure}

Next, to study the spectral variation in the intermediate state, 
we made 1-day averaged MAXI/GSC and \swift/BAT spectra in Phase B 
and applied the same model as in Sec.~\ref{subsec:spec_phase}: 
the {\tt TBabs*simpl*diskbb} model. The model successfully 
reproduced all the data. Figure~\ref{fig:spec_1d} shows 
two representative spectra at different X-ray fluxes and their 
best-fit models, and Figure~\ref{fig:trend_fitpars} presents 
the time variation of the 
best-fit parameters. At the brightest phase in soft X-rays 
(around MJD 59350; hereafter we call Phase B-0), 
the spectrum was dominated by the 
thermal component below 10 keV and the hard tail had a 
photon index of $\sim$ 2.0, 
while a steep power-law shaped spectrum with a photon index 
of 2.3--2.7 was seen at the other time periods in Phase B. 

\begin{figure}[ht!]
\plotone{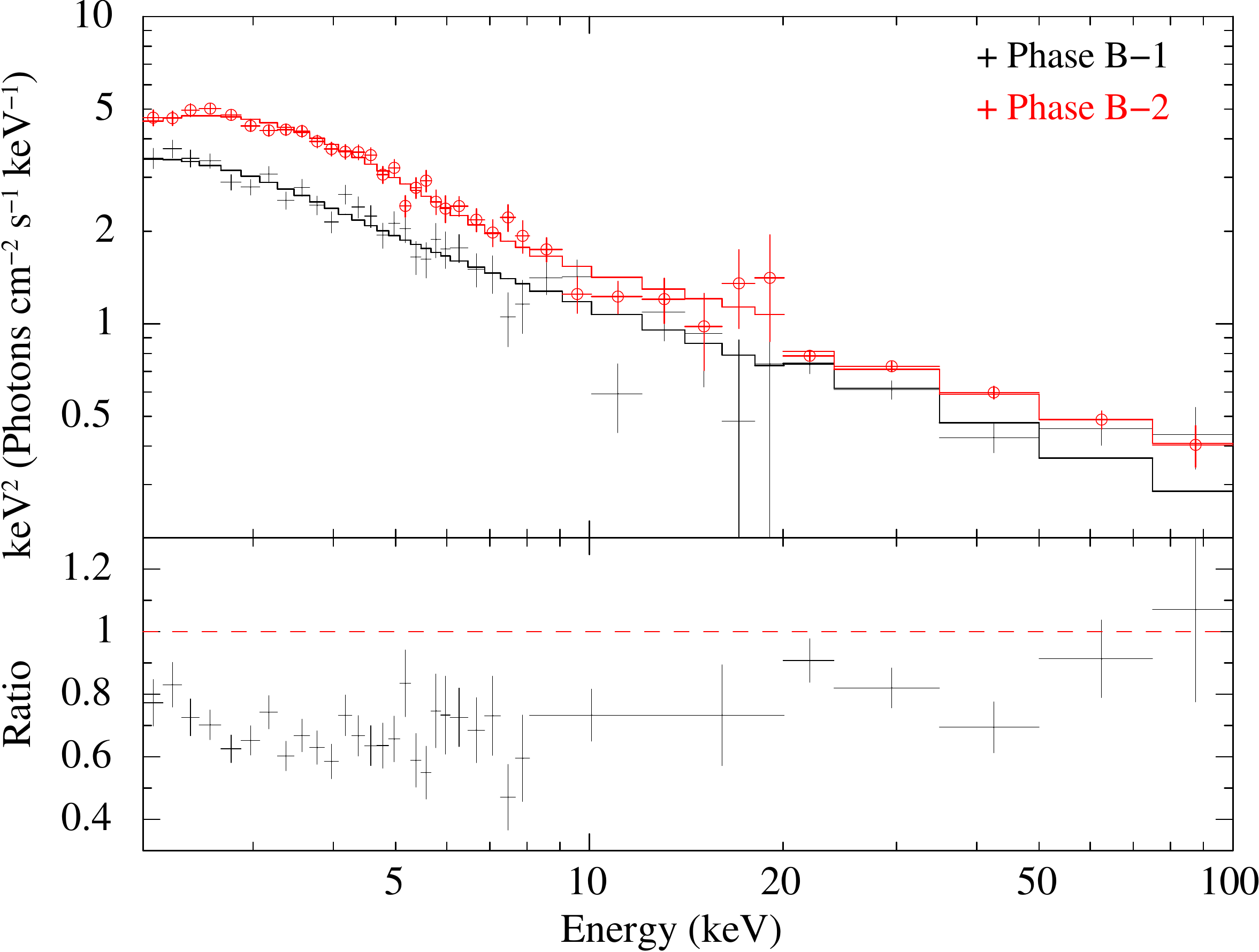}
\caption{(Top) Time-averaged unfolded spectra in 
Phase B-1 (black crosses) and B-2 (red open circles) 
and their best-fit models. (Bottom) Ratio of the {\tt folded} 
spectrum in Phase B-1 with respect to that in Phase B-2. 
The data points in 8--20 keV are binned further for visual 
purposes. 
\label{fig:spec_drop}}
\end{figure}

To investigate the cause of the flux drop seen in Fig.~\ref{fig:LC_4orb}, 
we also made time-averaged spectrum in MJD 59355.5--59358.0 (during 
the deepest flux drop; hereafter we call Phase B-1) and in
MJD 59358.5--59362.0 (after the drop; Phase B-2), and applied 
the same spectral model as above. 
The two spectra and their best-fit model are shown in 
Figure~\ref{fig:spec_drop}, and the best-fit 
parameters are given in Table~\ref{tab:fit}. We found that 
$\Tin$ was lower in the Phase B-1 than in Phase B-2. 
By contrast, $\rin$ and $\fscat$ favored larger values 
in Phase B-1, although they were marginally consistent 
each other between the two phases when the 90\% uncertainties 
were considered. 

The bottom panel in Figure~\ref{fig:spec_drop} shows the ratio of 
the raw spectra folded by the instrumental responses.  
The Phase B-1 spectrum has a slightly larger soft X-ray 
fraction below 3 keV than the Phase B-2 spectrum, which is 
consistent with the change in $\Tin$. 
We note that this flux drop cannot be explained by the 
increase in the absorption column density alone. In such case, 
softer X-rays would be more strongly absorbed and thus the 
spectrum would become significantly harder in the flux drop. 
To test if the absorption can explain at least some fraction of 
the flux drop at low energies, we fit the spectra with the same model 
above, allowing $\nh$ to vary. The fit only gave upper limits, 
$\nh \sim 4 \times 10^{21}$ cm$^{-2}$, for both spectra, but 
this value suggests that the absorption was not strongly 
enhanced in the flux drop.

\section{Discussion}

\subsection{Long-term Evolution and Spectral States}
We have performed long-term X-ray monitoring of the new Galactic 
black hole candidate \srcname~over 5 months since its discovery, 
using the MAXI/GSC and \swift/BAT. 
The long-term light curves were characterized by a rapid rise 
and slow decay, and the spectral softening was observed at the 
brightest phase (Fig.~\ref{fig:LC_longterm}). 
In the hardness intensity diagram, the source drew a 
counter-clockwise path (Fig.~\ref{fig:HID}) 
like other transient BHXBs \citep[e.g.,][]{miy95}. 
Recently, the soft-to-hard transition was reported, from 
NICER observations, to have occurred between MJD 59500 and 
59506 \citep{ste21}.
The low statistics of the MAXI/GSC data hampered 
the determination of the exact time of the transition, 
but the hardness intensity diagram  
suggests that the luminosity of the soft-to-hard 
transition was $\sim 1$ order of magnitude 
lower than that of the opposite transition. 

The combination of the two instruments enabled us to 
study the broad-band X-ray spectrum in 2--200 keV and its 
evolution during the outburst. At the initial rise of the 
soft X-ray flux (Phase A), the source showed a typical low/hard
state spectrum. The spectrum was well reproduced 
with a thermal Comptonization with $\Gamma \sim 1.7$ 
and $T_\mathrm{e} \sim 30$ keV, and the direct disk 
emission component was not observed. At the highest 
flux phase (Phase B), the soft X-ray fraction 
increased and the time-averaged spectrum was characterized 
by a steep power-law model with $\Gamma \sim 2.5$ consistent 
with the spectral profile in the intermediate or very high  
state. Although previous studies suggest disk truncation 
in this state \citep{tam12,hor14}, we did not detect 
a significant increase in $\rin$ likely 
due to the insufficient quality of the data. We note that 
the $\rin$ values obtained from the {\tt simpl*diskbb} model 
include the contributions of the scattered disk photons.

Due to the data gap of the MAXI/GSC, it is unclear exactly when 
is the onset of the transition, but most likely it occurred 
around the middle of the data gap of the MAXI/GSC;  
NuSTAR observed a typical low/hard state spectrum on 
MJD 59339--59340 \citep{xu21} while on MJD 59345--59346, 
AstroSAT detected a significant disk blackbody component with an 
inner disk temperature of $\sim 0.6$ keV, suggesting that the 
source was already in the intermediate state \citep{jan21}.
We have investigated the 1-day averaged \swift/BAT spectra 
in the data gap of MAXI, but detected no significant spectral 
variation due to low statistics. 

At the decaying phase after the intermediate state (Phase C--F), 
the spectrum was described by dominant multi-color disk blackbody 
emission in the soft X-ray band and a weak power-law tail
with $\Gamma \sim 2.0$ in the hard X-ray band. These spectral 
properties are consistent with the high/soft 
state. Applying the disk blackbody and its non-thermal
Comptonization model, we obtained the inner disk temperature 
$\Tin \sim$ 0.5--1 keV and the scattering fraction 
$F_\mathrm{scat} < $ 10\%, in agreement with typical BHXBs 
in the high/soft state \citep{mcc06}. Unlike the case 
of neutron star low mass X-ray binaries, no additional blackbody 
component was required in this period, supporting the black hole 
nature of the source. The idea that the source contains a 
black hole is consistent with the fact that no coherent pulsation 
has been detected so far \citep[e.g.,][]{xu21}. We note that 
search for pulsation using the MAXI/GSC data is difficult, because 
of the insufficient statistics and the very large time gap 
between each scan \citep[MAXI observes a source for only a few minutes 
in a 92 minute orbit; see][]{sug11}. Systematic search for pulsation 
of \srcname~is beyond our scope and we leave it as a future work. 
We have also searched the MAXI 2--10 keV light curve over the entire
outburst period but found no X-ray bursts. This also indirectly supports 
the black hole nature.

\subsection{Variation in the Intermediate State}
\label{sec:discussion_VHS}
Significant spectral variation on timescales of 
$\sim$day was observed in the intermediate state (Phase B). 
Around Phase B-0, which corresponds to the flux peak in the 
soft X-ray band, the spectrum was characterized by a dominant 
thermal disk component and a relatively weak hard tail 
with $\Gamma \sim 2.0$, 
reminiscent of the high/soft state spectrum. 
At lower luminosities, the source exhibited 
a steep power-law shaped spectrum with 
$\Gamma \sim 2.5$, consistent with the 
very high state spectrum, which is normally seen 
at higher luminosity than that in the high/soft 
state. Given that the steep power-law spectrum 
usually seen at sub-Eddington luminosities
(0.2--1 $\ledd$, where $\ledd$ is the Eddington 
luminosity), the source luminocity may have exceeded 
$\ledd$ at Phase B-0 and made the 
transition in the inner disk region 
from the standard disk to the slim disk. 
Using the best-fit {\tt simpl*diskbb} mdoel, we 
estimated the unabsorbed 0.01--100 keV flux at this phase 
of $\sim 2.4 \times 10^{-8}$ erg cm$^{-2}$ s$^{-1}$. 
If the source had $\sim 1 \ledd$ at this epoch, 
the unabsorbed 0.01--100 keV luminosity in Phase F 
is estimated to be $\sim 0.1 \ledd$. This is higher 
than the typical transition luminosity from the high/soft 
state to the low/hard state, $\sim$1\%--4\% $\ledd$
\citep{mac03, vah19}, and therefore suggests that 
the transition to the low/hard state should occur 
after Phase F. Indeed, a high/soft state spectrum was 
observed in Phase F and \citet{ste21} reported that the 
transition took place after Phase F.

In addition, we found a sudden flux drop with 
a duration of a few days, when the steep 
power-law spectrum was observed. Remarkably, 
the hardness ratio between the 4--10 keV and 
2--4 keV fluxes remained almost constant 
in this period (Fig.~\ref{fig:LC_4orb}). 
Comparing the spectra during and after 
the flux drop in detail, we found that the fraction 
of the flux around $\sim 2$ keV is slightly 
enhanced during the dip. 
Application of the {\tt simpl*diskbb} model yielded 
a smaller inner disk temperature and marginally 
larger inner disk radius and scattering fraction 
during the flux drop than after the drop. 

What produced this variation? One possibility is that 
the standard disk was truncated outside the innermost 
stable circular orbit (ISCO), and the transition 
from the standard disk to 
a strongly Comptonized accretion flow 
was taking place in the innermost region of the remaining 
disk \citep{tam12, hor14}.  
Such a transition, however, would progress as the 
mass accretion rate goes up. This contradicts 
what we observed: the disk receded as X-ray 
flux decreased. Another possibility would be that 
the mass accretion rate rapidly increased. 
In this case, the inner disk region may  
become geometrically thicker and reduce the apparent 
flux by shielding a part of the disk, 
although it is unclear whether the spectral shape can 
be kept almost constant. If the mass accretion rate 
goes up beyond the Eddington limit, a massive outflow 
can be launched in the inner disk region. If such  
outflow was really launched and was Compton thick 
and almost completely ionized, strong scattering could 
reduce the apparent X-ray flux without a great change in 
the spectral shape. A candidate of the 
Compton-thick wind was actually observed in GRO J1655$-$40
at similar luminosities \citep{shi16,nei16}. 

We note, however, that the very high state spectra are 
more complex than the {\tt simpl*diskbb} model 
\citep[e.g.,][]{gie03, tam12, hor14} 
and the spectral parameters that we estimated 
are taken with caution. Application of 
more realistic models to higher quality broad-band 
spectra would be required to investigate the details 
of the accretion flow structure at this epoch.

\subsection{Constraint on the Black Hole Mass} 

Previous studies suggested that the standard 
disk extends stably down to the ISCO in the high/soft 
state. Since the ISCO depends on the mass of the central 
compact object, we can constrain the black hole mass of 
\srcname~from the inner disk radius estimated in the 
high/soft state. In our spectral analysis, we obtained 
$\rin = 44 \pm 3$ ($\cos i/\cos 70^\circ$)$^{-1/2}$
($D$/8~kpc) km, from the normalization of {\tt diskbb}. 
Considering the correction factor of the boundary 
condition at the inner disk edge and the color-temperature 
correction factor ($K$ in total), 
the actual inner radius was estimated to 
be $\Rin = K \rin = 52 \pm 4$ ($\cos i/\cos 70^\circ$)$^{-1/2}$ ($D$/8~kpc) km. Here we adopted $K = 1.19$, assuming a color-temperature 
correction factor ($f_{\mathrm col}$) of 1.7 \citep{shi95}
and a torque-free 
inner boundary \citep{kub98}. Assuming a Schwarzchild 
black hole, we obtain a black hole mass of 
$\mbh = 5.8 \pm 0.4$ ($\cos i/\cos 70^\circ$)$^{-1/2}$ 
($D$/8~kpc) $M_\sun$. 

In the above discussion, we only considered the 
90\% error of $\rin$ obtained in the spectral fit, but 
the color-temperature correction factor could be 
an additional source of uncertainty. \citet{dav05} 
obtained $f_\mathrm{col} = 1.4-1.7$ from the calculation 
of the disk spectrum considering radiation transfer 
in the disk atmosphere and comparison with observations. 
If we adopt the lowest value, 1.4, $K$ is reduced by 
$\sim 30$\% and thereby $\mbh$ decreases by 
the same factor. Another uncertainty could be posed 
by the relativistic effects, which is not considered in 
the {\tt diskbb} model. To investigate the effects, 
we fit the Phase C--F spectra simultaneously in the 
same manner as in Section~\ref{subsec:spec_phase}, 
but replacing {\tt diskbb} to the relativistic disk 
emission model {\tt kerrbb} \citep{li05}. 
In this fit, we allowed the mass accretion rate 
and $\mbh$ to vary and fixed all the other parameters 
of {\tt kerrbb}: 
$f_\mathrm{col} = 1.7$, $i = 70^\circ$ or $85^\circ$, 
$D=8$ kpc, $a_* = 0$, where $a_*$ is the dimensionless 
spin parameter of the black hole defined as $Jc/GM^2$ 
($J$ is the angular momentum of the black hole). 
We assumed a torque-free inner boundary and considered 
the self-irradiation effect but ignored the limb-darkening 
effect. From this model, $\mbh$ was estimated to be 
$6.1^{+0.4}_{-0.3}$ and $10.5^{+0.7}_{-0.5}$ 
for $i = 70^\circ$ and $85^\circ$, respectively. 
These values, although having large errors, are 
consistent with the values that we obtained  
from the {\tt diskbb} model, and hence the relativistic 
effects can be considered to be weak in a non-rotating 
black hole.

In the case of a Kerr black hole with a prograde spin, 
the relativistic effects can be stronger because the ISCO 
radius decreases with the spin, and the observed 
disk spectrum can significantly deviate from 
what {\tt diskbb} predicts. Generally, the black hole mass 
estimated from the relativistic disk model increases as the 
spin increases and the ISCO radius decreases, 
but the exact value depends on the black hole spin 
and the inclination angle in a complicated way 
\citep[e.g.,][]{shi11,wan18}. 
Estimating the black hole mass and spin and 
the inclination simultaneously is difficult for 
the MAXI/GSC data, due to insufficient statistics 
and lack of data in the soft X-ray band below 2 keV. 
Application of the relativistic disk model to a 
better quality soft X-ray spectra is left as a future work. 

\begin{figure}[th!]
\plotone{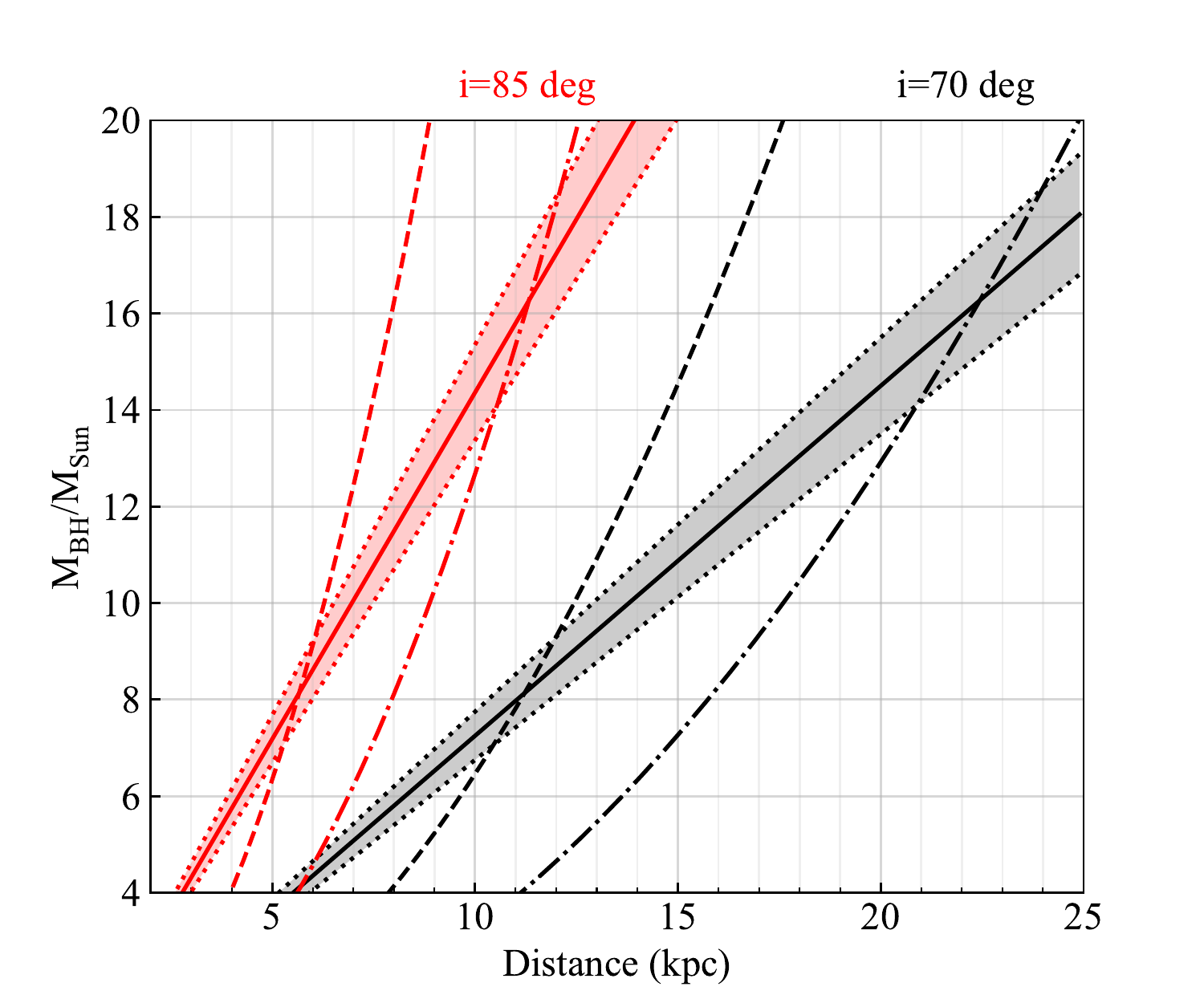}
\caption{Constraints on the black hole mass and distance of 
\srcname~for an inclination of $70^\circ$ (black) and 
$85^\circ$ (red), where a Schwarzchild black hole is assumed. 
The solid lines are derived  
from the inner disk radius in the high/soft state, and 
the shadowed region are their 90\% error ranges. 
The dashed and dash-dotted lines show the relations obtained 
when the peak luminosity is 0.5 and 1.0 $\ledd$, respectively. 
\label{fig:M_D}}\end{figure}

In Figure~\ref{fig:M_D} we plotted the 
$\mbh$ versus $D$ relation obtained above.  
While the inclination angle of \srcname~is likely 
to be $i \gtrsim 70^\circ$ because of 
the presence of winds and absorption dips,  
the distance have not been constrained so far. 
Considering the luminosity at the soft X-ray peak 
(MJD~59350) was likely close to 
the Eddington luminosity (see 
Section~\ref{sec:discussion_VHS}), we can get 
another constraint on $\mbh$ and $D$. 
This is also plotted in Fig.~\ref{fig:M_D}. Here, 
the unabsorbed 0.01--100 keV flux at the soft X-ray 
peak, $2.4 \times 10^{-8}$ erg cm$^{-2}$ s$^{-1}$, 
was adopted and was converted to the luminosity 
$L_\mathrm{peak}$ by assuming emission 
from a geometrically thin disk surface.

Combination of the $D$ versus $\mbh$ relations 
from the ISCO radius and peak luminosity favors  
moderate or large distances, $\gtrsim 11$ kpc for 
$i=70^\circ$, and $\gtrsim 6$ kpc for $i=85^\circ$ 
when $L_\mathrm{peak} > 0.5 \ledd$ is assumed.
The source is therefore likely located near 
the galactic center or farther. This is in 
agreement with the fact that the absorption 
column density estimated with a NICER spectrum is consistent 
with the total Galactic column $\nh \sim 3 \times  10^{21}$ cm$^{-2}$ 
obtained from the ftool {\tt nh}, which calculates the hydrogen 
column density based on the 2D HI4PI map, a full-sky HI survey \citep{hi416}. 
A relatively large black hole mass $\gtrsim 8 \msolar$ was 
obtained in either case of the inclination angles. 
 
\section{Summary}
Using the MAXI/GSC and \swift/BAT data, 
we have studied the X-ray spectral evolution 
of the new Galactic black hole candidate \srcname.
The source showed the state transition from 
the low/hard state to the high/soft state via 
the intermediate state. The flux variation on 
timescale of $\sim 1$ day was detected in the 
intermediate state, which could be interpreted as 
a rapid change in the mass accretion rate 
(and possibly launch of a Compton-thick outflow)
at the transition between the standard disk 
and the slim disk. Using the inner disk 
radius in the high/soft state and 
the peak luminosity, we estimated a black hole mass 
and distance of \srcname~to be $\sim 8 \msolar$ 
and $\gtrsim 6$ kpc (where a non-spinning black hole, 
an inclination of $\gtrsim 70^\circ$, and the 
peak luminosity of $\gtrsim$ 0.5 times the Eddington 
luminosity are assumed), suggesting that the source is 
a black hole X-ray binary located near or farther 
than the galactic center region. 

\acknowledgments
This work made use of 
MAXI data provided by RIKEN, JAXA, and the MAXI team, 
and from Data ARchives and Transmission System (DARTS)
at ISAS/JAXA, and of \swift~public data 
from the \swift~data archive. 
Part of this work was financially supported 
by Grants-in-Aid for Scientific Research 19K14762 (MS) 
from the Ministry of Education, Culture, Sports, 
Science and Technology (MEXT) of Japan.



\facilities{MAXI~(GSC), \swift~(BAT), NICER~(XTI)}


\software{
XSPEC (v12.11.1; \citealt{arn96}), HEAsoft (v6.28; \citealt{heasoft}))
          }

\vspace{25mm}


\bibliography{maxij1803_ms}{}
\bibliographystyle{aasjournal}


\appendix
\section{MAXI/GSC foloded spectra in the individual phases}
\label{appsec:MAXIspec_folded}
\begin{figure*}[ht!]
\epsscale{1.15}
\plotone{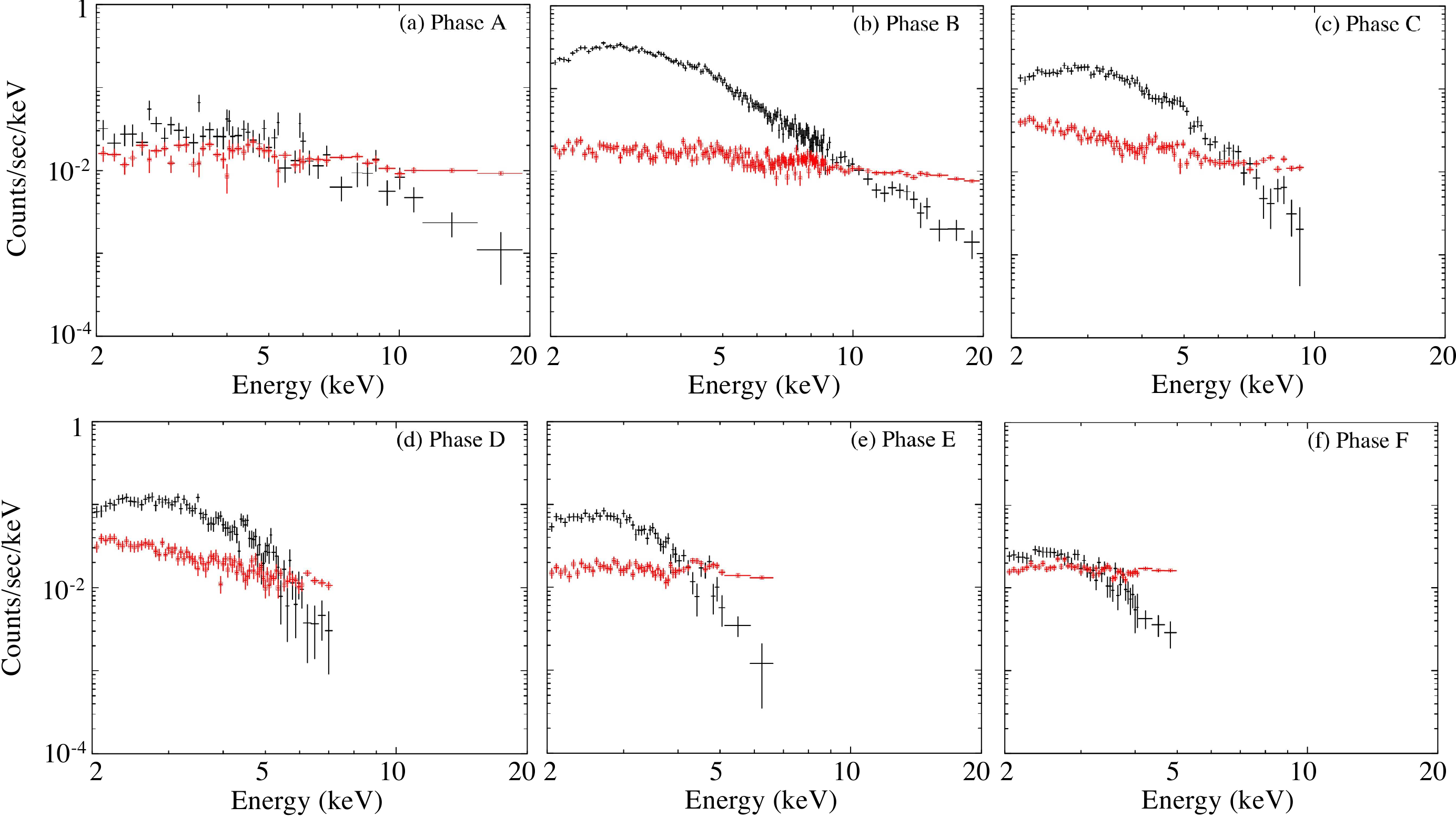}
\caption{The MAXI/GSC response-folded, background-subtracted 
spectrum of \srcname~in Phase A--F. The background spectra are 
shown with red open squares. 
\label{fig:MAXIspec_folded}}
\end{figure*}

\end{document}